\documentclass[prl,twocolumn,superscriptaddress,showpacs]{revtex4}

\usepackage{graphicx}%

\newcommand{\AAs}{\AA\hspace{1ex}} 

\begin{document}
\bibliographystyle{apsrev}


\title{Using dynamically scattered electrons for 3-dimensional 
potential reconstruction}

\author{Christoph T. Koch}
\email[]{koch@mf.mpg.de}

\affiliation{Max Planck Institute for Metals Research, 
Heisenbergstr. 3, 70569 Stuttgart, Germany }

\date{\today}

\begin{abstract} 

Three-dimensional charge density maps computed by first-principles methods 
provide information about atom positions and the bonds between them, data  
which is particularly valuable when trying to understand the properties of 
point defects, dislocations, and interfaces.
This letter presents a method by which 3-dimensional maps 
of the electrostatic potential, related to the charge density map by 
the Poisson equation, can be obtained experimentally at 1 \AAs resolution 
or better.
This method requires data acquired by holographic transmission electron 
microscopy (TEM) methods such as off-axis electron holography or focal 
series reconstruction for different directions of the incident electron 
beam.
The reconstruction of the 3-dimensional 
electrostatic (and absorptive) potential is achieved by making use of 
changes in the dynamical scattering within the sample as the direction of 
the incident beam varies.   

\end{abstract} 


\maketitle

Extended defects such as dislocations, interfaces, in particular solid 
liquid interfaces, pose a particular challenge to atomic scale 
computational methods.   
While their complexity normally requires super-cells too large for 
ab-initio
methods, the long-range nature of forces determining their properties  
(e.g. strain fields, Coulomb and van der Waals [dispersion] forces) makes 
the construction  of reliable yet efficient interatomic potentials 
required for molecular dynamics (MD) and related computational methods an 
extremely complicated task. 
Being able to perform experiments which directly map
the 3-dimensional local electrostatic potential would provide a wealth of 
information without the need to do any simulations at all and, if 
computations are needed to extract information not provided by the 
potential map or, in the case of MD, it would allow verification of 
interatomic potentials by direct comparison between experiment and 
simulation for the very (defect) structure under investigation.
For single crystals of small unit cell it has recently been demonstrated 
that it is possible to map the bond charge
distribution by fitting Fourier coefficients of the 
crystal potential to convergent beam electron diffraction 
(CBED) data \cite{Zuo99}.

TEM images, in the context of focal series reconstruction 
(inline  holography) over a large defocus range \cite{Koch08FRWR}, are 
very sensitive to
relative atomic positions and variations in the electrostatic potential.   
Although claiming to be able to provide data on the 3-dimensional 
distribution
of atoms this is not true for destructive methods such as 3D atom probe 
\cite{Blavette93}
because  the atoms being "imaged" are not extracted from their original  
environment, but from the sample surface.  

The enormous advantages of TEM  
imaging come at the price of the lack of a general direct interpretability  
at atomic resolution.  Although modern electron optics has been able to  
remove many of the artefacts caused by aberrations of the imaging system,  
it cannot remove artefacts produced by the multiple (or dynamical) 
scattering  process itself.  This is the main reason why electron 
tomography  has never been applied at atomic resolution, with the 
exception  of extremely small nanostructures consisting of light atoms, 
for which the authors felt that they could neglect dynamical electron 
scattering events \cite{BarSadan08}. 
Attempts to correct for artefacts (i.e. features which cannot be 
interpreted directly in terms of atomic structure) in the exit face wave 
function
produced by dynamical scattering effects have, in the case of thin or 
non-crystalline samples,
at most been able to reconstruct an average projected potential 
\cite{Gribulek91,Beeching93,Scheerschmidt98, Lentzen00, Allen01_dyn}.

The method presented in this letter directly reconstructs the local 
complex scattering
potential on a 3-dimensional grid, the real part of which is the 
electrostatic potential,
by separating the multiple scattering paths between potential voxels.

\begin{figure}
\begin{center}
\includegraphics[width=0.48\textwidth]{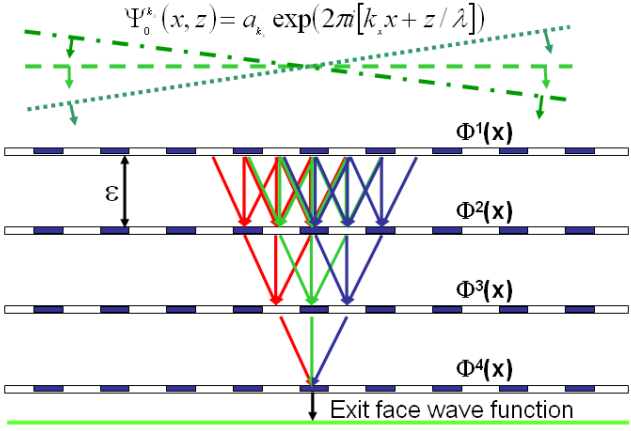}%
\end{center}
\caption{(color online) Diagram of real-space multislice algorithm.  The 
scattering of the incident electron beam described by a plane wave 
is approximated by a finite number of scattering events at equidistant 
layers partitioning the sample in \(z\)-direction.  As a consequence, 
the signal in the exit face wave function is non-local, including 
contributions from a number of scattering paths, the relative phase of 
which can be varied by changing the illumination tilt angle 
\(\theta_x = \sin^{-1} k_x \lambda\).} 
\label{Fig_multislice} 
\end{figure}


The following discussion will, in order to keep the equations  
readable, only consider the reconstruction of a 2-dimensional structure 
from
a series of 1-dimensional images.   The extension to 3 dimensions is 
straight forward, requiring only small changes in the presented 
formalism.

The proposed reconstruction method is based on the
real-space variant \cite{vanDyck80} of the multislice algorithm \cite{Cowley57} 
for solving the relativistically corrected Schr{\"o}dinger equation 
describing
the propagation of a fast electron through the specimen potential
(see figure \ref{Fig_multislice} for an illustration).
In a first step
the 
2-dimensional scattering potential (remember that this 
discussion can easily be extended to 3 dimensions) 
is divided
into a set of 
\(N\) discrete horizontal slices
of thickness \(\epsilon\) (the optical axis of the microscope is assumed 
to be vertical, the fast
beam electrons travelling down the microscope column) and the 
potential within a slice of index \(m\) is projected into a 
1-dimensional layer of potential 
\begin{eqnarray}
\nonumber 
V^{(m)}(x) = \int_{\epsilon (m-1/2)}^{\epsilon (m+1/2)}
V(x,z) dz.
\end{eqnarray}

The electron propagation is then approximated by multiplication of the 
incident wave function by a phase grating \(\Phi^{(m)}(x) = \exp[i \sigma 
V^{(m)}(x)])\) (\(\sigma=2 \pi \left(m_e
c^2+E_0\right) / \left[\lambda E_0 \left(2 m_e c^2+E_0\right) \right]\) 
being the electron-potential interaction constant) at the position of each 
slice and Fresnel
propagation between the slices.  We will assume an incident plane wave 
\(A \exp(2 \pi i \vec{k} \cdot \vec{r}) = A \exp[2 \pi i (k_x x + k_z 
z)]\) arriving at the sample at an angle \(\theta = \sin^{-1}(k_x/k_z)\). 
Since holographic experiments can measure relative phases but not the 
absolute phase of an electron wave functions we will define \(z=0\) in the 
plane of the exit face wave function, fixing the 
z-dependence of the wave function at the entrance surface.  Combining this 
z-dependent phase factor with the scaling factor \(A\) we obtain 
\(\Psi^{(k_x)}(x) = A^{(k_x)} \exp[2 \pi i k_x x ]\) at the entrance 
surface of the specimen. 
The 
electron
wave function at the exit surface can then be obtained by the real-space 
multislice formalism \cite{vanDyck80} as
\begin{eqnarray}
\nonumber 
\Psi^{(k_x)} (x) &=& \Phi^{(N)}(x) \cdot e^{\epsilon \lambda
\left(
\frac{i \Delta_{x}}{4 \pi}-k_x \nabla_x\right)} \cdots  \\
\nonumber 
& & \cdot \Phi^{(2)}(x) \cdot 
e^{\epsilon \lambda\left(\frac{i \Delta_{x}}{4 \pi}-k_x \nabla_x\right)}
\Phi^{(1)}(x) \\
& &  \cdot A^{(k_x)} \exp[2 \pi i k_x x]
\label{multislice}
\end{eqnarray}
Here \(\lambda = 1/k_z = h c/\sqrt{E_0 \left(2m_e+E_0\right)}\) is 
the
electron wavelength, \(E_0\) the accelerating voltage, \(h\) Planck's 
constant, \(c\) the speed of light, \(m_e\) the mass of an electron, 
\(\Delta_{x}\)  the 
Laplace operator and \(\nabla_x\) the gradient in the \(x\)-direction.

Implementing the multislice algorithm on a computer requires a  
discrete grid in the lateral dimension in addition to the discrete 
slices in the \(z\)-direction.
 Keeping the lateral sampling distance \(\delta_x\)  constant we 
may simplify the notation by 
defining a dimensionless imaginary parameter \(\gamma_{k_x} = 2 \pi i 
\delta_x k_x \) and
replacing \(x\) by \(j_{x} \delta_x\),
\(\Psi(x)\) by \(\Psi_{j_{x}}\), \(V^{(m)}(x)\) by \(V^{(m)}_{j_{x}}\), 
\(\Phi^{(m)}(x)\) by \(\Phi^{(m)}_{j_{x}}\), and 
\(\exp[2 \pi i k_x x]\) by \(\exp[\gamma_{k_x} j_{x}]\). 
This makes also the translation of 
the following equations into a computer program a bit more apparent.


Expanding the Fresnel propagation operator by using the relations
\begin{eqnarray}
\nonumber 
\Delta_{x} \Psi(x) & = & \frac{\Psi(x+\delta_x) - 2\Psi(x) + 
\Psi(x-\delta_x)}{\delta_x^2} \quad \textrm{ and}\\
\nonumber 
\nabla_x \Psi(x) & = & \frac{\Psi(x+\delta_x) - \Psi(x-\delta_x)}{2 
\delta_x} 
\end{eqnarray}
we obtain 
\begin{eqnarray}
\nonumber 
& & \exp \left[\frac{i \epsilon \lambda}{4 \pi} 
\left( \Delta_{x} + \gamma_{k_x} \nabla_x \right)\right] \Psi(x) \\
\nonumber 
&=& \left\{1+\frac{i \epsilon \lambda}{4 \pi}
\left( \Delta_{x} + \gamma_{k_x} \nabla_x \right) \right. \\
\nonumber 
& & + \left. \frac{1}{2} \left[ \frac{i \epsilon \lambda}{4 \pi}
\left( \Delta_{x} + \gamma_{k_x} \nabla_x \right) \right]^2
+ \ldots \right\} \Psi(x) 
\\ \nonumber 
&=& \Psi(x) + \beta \left[(1+\gamma_{k_x})\Psi_{j_x+1} - 2\Psi_{j_x} + 
(1-\gamma_{k_x})\Psi_{j_x-1} \right]  \\
\nonumber 
& & + \frac{\beta^2}{2} \left[(1+\gamma_{k_x})^2\Psi_{j_x+2} - 
4(1+\gamma_{k_x})\Psi_{j_x+1}  \right.  \\
\nonumber 
& & \left. + 
(6-2\gamma_{k_x}^2)\Psi_{j_x} - 4(1-\gamma_{k_x})\Psi_{j_x-1} \right. \\
\label{Laplace_expansion} 
& & \left.
+ (1-\gamma_{k_x})^2\Psi_{j_x-2} 
\right] + O\left(\beta^3 \right)
\end{eqnarray}
in orders of approximation of the 
Ewald sphere curvature given by the parameter \(\beta = i \epsilon \lambda 
 / \left(4 \pi \delta_x^2\right)\).  
The sampling distance \(\delta_x\)
is typically quite a bit smaller than the image resolution defined by
aberrations of the electron optics, the stability of the 
microscope, and the size of the objective aperture 
(e.g. \(0.1 \ldots 0.5\)\AAs for HRTEM at 1 \AAs resolution).

By chosing the appropriate real-space sampling, slice thickness and 
electron accelerating voltage one can force  
\(\beta\) to have a fairly small value.
Plugging the expansion of the Fresnel propagator (\ref{Laplace_expansion}) 
into (\ref{multislice}) we can expand the expression for the exit face 
wave
function in orders of \(\beta\).  The first order expansion 
neglecting, for the moment, terms of order \(\beta^2\) and higher looks 
as follows: 
\begin{eqnarray}
\nonumber 
\Psi^{(k_x)}_{j_{x}} &=& A^{(k_x)} e^{\gamma_{k_x} j_{x}} 
\Phi^{(N)}_{j_{x}}  \left\{ \prod_{m=1}^{N-1} 
\Phi^{(m)}_{j_{x}} \right. \\
\nonumber 
& & \left. +\beta \left[ e^{\gamma_{k_x}} (1+\gamma_{k_x})
\sum_{k=1}^{N-1} 
\prod_{m = 1}^k \Phi^{(m)}_{j_{x}+1} 
\prod_{m = k+1}^{N-1} \Phi^{(m)}_{j_{x}}  \right. \right. \\
\nonumber 
& &  \left.  \left.
+e^{-\gamma_{k_x}} (1-\gamma_{k_x})
\sum_{k=1}^{N-1} 
\prod_{m = 1}^k \Phi^{(m)}_{j_{x}-1}
\prod_{m = k+1}^{N-1} \Phi^{(m)}_{j_{x}} \right. \right. \\
\label{Psi_1}
&& \left.  \left. -2 (N-1) \prod_{m=1}^{N-1} \Phi^{(m)}_{j_{x}} \right] + 
O(\beta^2) \right\} 
\end{eqnarray}
Note here, that the first term, the zeroth order expansion is independent
of \(\beta\) and is just the well-known phase object approximation which 
includes multiple scattering to \(N^{th}\) order but neglects effects 
due to curvature and tilt of the Ewald sphere.

\begin{figure}
\begin{center}
\includegraphics[width=0.48\textwidth]{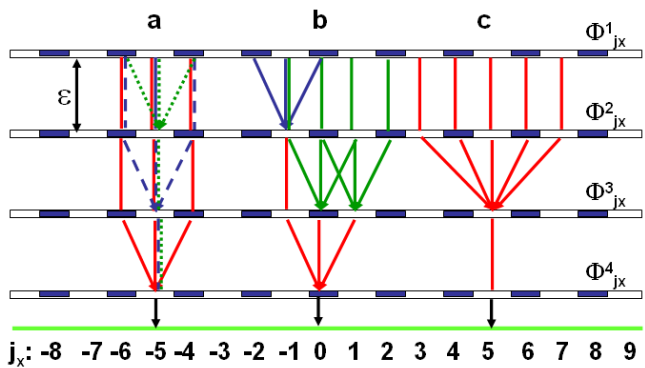}%
\end{center}
\caption{(color online) Diagram of first (A) and second order (B,C) 
approximation
of the Ewald sphere curvature effects in the real-space multislice 
algorithm for \(k_x = 0\).  The \(\Delta_x\) operator is depicted by a 
node merging
3 paths, and the \(\Delta_x^2\) operator by a node merging 5 paths. 
Multiplying the scattering potential along the the read solid, blue dashed 
and green dotted paths in A will produce all the possible monomials that
are linear in \(\beta\), i.e. which involve only a single \(\Delta_x\) 
operator.  Monomials resulting from paths shown in B and C (only one 
possible path is shown in each) are proportional to \(\beta^2\) and 
involve either 2 successive \(\Delta_x\) operations, or a single 
\(\Delta_x^2\) operation.} 
\label{scatDiag} 
\end{figure}

Figure \ref{scatDiag}a illustrates expression (\ref{Psi_1}) graphically 
for an object partitioned into 4 layers.   Examples of the 2 possible 
scattering paths that scale as \(\beta^2\) are shown in figure 
\ref{scatDiag}b and \ref{scatDiag}c.  While the conventional multislice 
algorithm \cite{Ishizuka77} iterating  between real and reciprocal space 
naturally includes all orders \(\beta^n\), in real-space algorithms they 
need to be included explicitly.  Since increasing \(n\) slows down the 
computation enormously, terms \(O(\beta^{n\geq 3})\) in the computation 
of a single Fresnel propagation are usually neglected.
An algorithm approximating the Fresnel propagation up to 
\(O(\beta^{n=1})\)
will only have contributions of \(O(\beta^{n\geq 2})\) to the final wave 
function of the type shown in figure \ref{scatDiag}b, i.e. multiple 
\(\Delta_x\) operations, but neglect the equally important contributions 
shown in figure \ref{scatDiag}c.  
 
The aim of this letter being the determination of 
the 3-dimensional scattering potential from the observed images we must 
consider the contributions to the exit face wave function in the order of 
significance, i.e. all terms up to a given order \(O(\beta^n)\).  Taking 
a closer look at figure \ref{scatDiag} makes clear that in order to 
reconstruct \(N\) slices one needs to expand the Fresnel propagation up to  
at least \(n = (N-1)/2\) (rounding up for even \(N\)).  The resulting 
system of polynomial equations of degree \(N\) is sparse and may be solved 
using globally convergent algorithms for solving multivariate polynomial 
sets of equations (see e.g. \cite{Sherali92} with an example application 
given in \cite{Koch08_LACBED}).


Holographic methods such as off-axis electron holography \cite{Moellenstedt56}, but also focal 
series reconstruction 
(see, for a very recent example, \cite{Koch08FRWR} and references to 
9 alternative algorithms therein)
are able to reconstruct the complex exit face wave function for each 
incident beam tilt.  
If, as is common practice, in the off-axis holographic reconstruction the 
side band
is properly centered, and in the focal series reconstruction the images 
are aligned by cross-correlation or similar methods,  neither of these 
methods will reconstruct the global phase shift 
\(\exp[\gamma_{k_x} j_x]\)
that is associated with the tilted illumination (the relative phase 
factors in the second and third term of (\ref{Psi_1}) remain, though).
We can therefore drop this term alltogether.
However, 
a reference point common to wave functions of different incident beam tilt 
must still be chosen in order to fix the complex coeffcients 
\(A^{(k_x)}\).  A vacuum region or another area of well-known scattering 
properties within the field of view would be ideal, but if no such 
reference point can be defined the \(A^{(k_x)}\) parameters can also be 
included in the non-linear reconstruction algorithm.  We will assume  
\(A^{(k_x)} = 1\) in the example presented below.

\begin{figure}
\begin{center}
\includegraphics[width=0.48\textwidth]{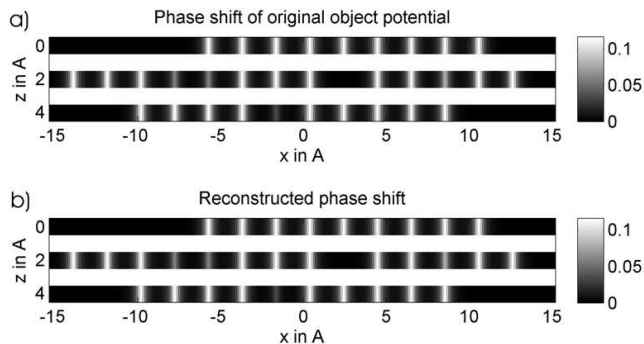}%
\end{center}
\caption{Phase shift \(\phi^{(m)}(x) = \sigma V^{(m)}(x)\) for a 3-layered 
''phantom'' structure (a) and the phase shift reconstructed from it 
using the linear approximation to (\ref{Psi_1}) (b)  
demonstrating a 3-dimensional reconstruction at
atomic resolution.  The scattering factor and lattice constant correspond 
to that of gold.  The ''defect'' sites have been created by simply scaling 
the scattering factor of gold by a number less than 1 
(individual defects from left to right: second row: 0.5, 0.5, 
0, 0.7, third row: 0.3).  Apsorption has been included by multiplying the 
potential \(V^{(m)}(x)\) by \((1+0.1i)\).
Simulation parameters: object potential grid: \(240 \times 3\)  pixels, 
\(\delta_x=0.128\)\AA, \(\epsilon=2.04\)\AA, \(E_0=60\)kV, \(\sigma = 
0.0011\) (V\AA)\(^{-1}\), \(\beta = 0.49i\),
5 different tilt angles of the incident beam: \(\theta=-1^{\circ}\) 
(\(\gamma_{k_x}
= -0.29i\)), \(\theta=-0.3^{\circ}\) (\(\gamma_{k_x} = -0.086i\)),
\(\theta=0^{\circ}\) (\(\gamma_{k_x} = 0i\)),
\(\theta=0.7^{\circ}\) (\(\gamma_{k_x} = 0.20i\)),
\(\theta=1^{\circ}\) (\(\gamma_{k_x} = 0.29i\)).
} 
\label{Fig_reconstruction} 
\end{figure}

For demonstration purposes, and in order to reveal the principles of the 
proposed methodology independent of the performance of a given polynomial 
equation
solver we approximate the system of polynomial equations (\ref{Psi_1}) by 
expanding the phase grating \(\Phi^{(m)}_{j_x}\)  in terms of the 
potential  assuming
the validity of the weak phase object approximation.
For a structure that has been split into \(N=3\) layers we 
approximate 
\begin{eqnarray}
\nonumber 
\Phi^{(3)}_{j_{x3}} \Phi^{(2)}_{j_{x2}} \Phi^{(1)}_{j_{x1}}
&=& e^{i \sigma V^{(3)}_{j_{x3}}}  e^{i \sigma V^{(2)}_{j_{x2}}} 
  e^{i \sigma V^{(1)}_{j_{x1}}} \\
\nonumber 
&\approx& 1+i \sigma V^{(3)}_{j_{x3}} + i \sigma V^{(2)}_{j_{x2}} 
  + i \sigma V^{(1)}_{j_{x1}}
\end{eqnarray}
converting (\ref{Psi_1}) into a linear system of equations, 
the solution of which should be straight forward.  It turns out the the 
matrix defining the resulting system of equations is slightly 
rank-deficient, because dropping all the non-linear terms in 
\(V^{(m)}_{j_x}\) introduces 2 additional degrees of freedom, the 
overall
phase relationship between the 3 layers.  Defining the relative potential 
of 1 column of pixels (e.g. in the example shown in figure 
\ref{Fig_reconstruction} I defined 3 additonal linear equation that 
force the potential to be zero (vacuum) for the right most pixel in each 
of the 3 layers) will remove this degree of freedom and produce, in the 
absence of noise, a unique reconstruction.
The system of linear equations has been solved using singular value 
decomposition.

Figure \ref{Fig_reconstruction} a shows the phase shift of a trial 
structure that has been sliced into 3 equidistant slices.  Scattering 
factors and length scales have been scaled to those of the gold 
(110) lattice, in order to mimic actual experimental conditions.  Since no 
noise has been included in this test, it is not surprising that  
reconstruction and original look identical.  The fairly low accelerating 
voltage of only 60kV causes \(\sigma\) (multiple scattering strength), 
\(\beta\) (Ewald sphere curvature)
and \(\beta \gamma_{k_x}\) (illumination tilt sensitivity) to be of large 
enough values to make a 3-dimensional atomic resolution
reconstruction of nano structures feasible, even in 
the presence of noise and for small beam tilt angles.

Some iterative tomographic reconstruction algorithms allow the specimen 
geometry to be used as a constraint, in principle being able to 
reconstruct
an object consisting of only \(N\) distinct layers to be reconstructed 
from
\(N\) projections only.  The angle between these \(N\) projections 
must, however, be quite large.
Although requiring the same minimum number of tilt angles,  
in contrast to (linear) tomography, where  the 3-dimensional 
information is introduced by projecting along different directions, 
making use of dynamical scattering as proposed here, implies a tilt-angle 
dependent (complex) weighting
factor for a set of monomials in the polynomial system of equations 
(\ref{Psi_1}).  These weighting factors depend strongly on the 
accelerating voltage and can therefore be tuned to the 
scattering strength of the material, available beam tilt range and 
desired resolution in the direction parallel to the electron beam.
Modern TEMs being able to achieve atomic 
resolution imaging at \(E_0 \leq 60 kV\) (e.g. \cite{Kisielowski08}) will 
therefore
be able to image the 3-dimensional potential distribution within 
nanostructures at atomic resolution without having to tilt the specimen.

It should be noted that exit wave reconstruction for several 
illumination tilt angles is only one of several methods for acquiring 
the experimental data required
for the 3D reconstruction described above.  Alternative methods include 
R{\"o}nchigram focal series recorded for different illumination or 
specimen shifts and, of course, holographic experiments at different 
specimen tilts.

Summarizing, a method for reconstructing the 3-dimensional electrostatic 
potential of a TEM sample has been proposed.
The method is based on a reformulation of the real-space multislice 
formalism for computing the multiple scattering of a fast electron within  
a TEM sample in terms of a polynomial set of equations, identifying and 
keeping the most significant terms. 
The solution of the resulting sparse multivariate polynomial system of 
equations can be found by applying polynomial equation global optimization 
algorithms available in the literature.
 
In the special case that the weak phase object approximation is valid the 
polynomial system of equation transforms into a linear system of equations 
which can be solved by standard methods.
Applying this simplification a 2-dimensional complex test 
object consisting of 3 layers has been reconstructed successfuly from 5 
1-dimensional electron wave functions simulated for illumination tilt 
angles ranging from \(-1^{\circ}\) to \(+1^{\circ}\).
The construction of an efficient global optimization algorithm 
specialized for
the particular type of systems of polynomial equations described in this 
letter is planned for the near future.

\begin{acknowledgments}
This work was supported by the European Commission under contract 
Nr. NMP3-CT-2005-013862 (INCEMS).
\end{acknowledgments}


\begin{thebibliography}{10}

\bibitem{Zuo99}
J.M. Zuo, M. Kim, M. O' Keeffe, J.C.H. Spence, Nature {\bf401}, 49 (1999)

\bibitem{Blavette93}
D. Blavette, B. Deconihout, A. Bostel, J.M. Sarrau, M. Bouet, A. Menand, 
Rev. Sc. Instr. {\bf64}, 2911  (1993)

\bibitem{BarSadan08}
M. Bar Sadan, Nano Letters  {\bf8}, 891 (2008)

\bibitem{Gribulek91}
M.A. Gribelyuk, Acta Cryst. A{\bf47}, 715 (1991)

\bibitem{Beeching93}
M.J. Beeching and A.E.C. Spargo, Ultramicroscopy {\bf52}, 243 (1993) 
243.

\bibitem{Scheerschmidt98}
K. Scheerschmidt, J. Microscopy {\bf190}, 238 (1998) 
243.

\bibitem{Lentzen00}
M. Lentzen and K. Urban, Acta Cryst. A{\bf56}, 235 (2000)

\bibitem{Allen01_dyn}
L.J. Allen, C.T. Koch, M.P. Oxley, and J.C.H. Spence, Acta Cryst. 
A{\bf57}, 473 (2001)


\bibitem{vanDyck80}
D. van Dyck, J. Microscopy {\bf119}, 114 (1980)

\bibitem{Cowley57}
J.M. Cowley and A.F. Moodie, Acta Cryst. {\bf10}, 609 (1957)



\bibitem{Moellenstedt56}
G. M{\"o}llenstedt and H. D{\"u}ker, Z. Phys. {\bf145}, 377, (1956)










\bibitem{Koch08FRWR}
C.T. Koch, Ultramicroscopy {\bf108}, 141 (2008)

\bibitem{Ishizuka77}
K. Ishizuka and N. Uyeda, Acta Cryst. A{\bf33}, 740 (1977)


\bibitem{Sherali92}
H. D. Sherali and C. H. Tuncbilek, Journal of Global Optimization {\bf2}, 
101-112 (1992)

\bibitem{Koch08_LACBED}
C. T. Koch, Phys. Pev. Lett. {\it submitted} (2008)


\bibitem{Kisielowski08}
C. Kisielowski et al., Microscopy and Microanalysis {\bf14}, 469 (2008)

\end{thebibliography}
\end{document}